\documentclass{article}
\usepackage{spconf,amsmath,graphicx}
\usepackage{subcaption}
\usepackage{subfig}
\title{AUDIO-BASED DISTRIBUTIONAL SEMANTIC MODELS \\[1mm] FOR MUSIC AUTO-TAGGING AND SIMILARITY MEASUREMENT}

\name{\begin{tabular}{c}Giannis Karamanolakis$^1$ \qquad Elias Iosif $^1$ \qquad Athanasia Zlatintsi$^1$\\ Aggelos Pikrakis$^2$\qquad Alexandros Potamianos$^1$\end{tabular}}

\address{$^1$School of Electrical \& Computer Engineering, National Technical University of Athens, Greece \\
    $^2$Department of Informatics, University of Piraeus, Greece\\
    \small \tt giannis.karamanolakis@gmail.com, iosife@central.ntua.gr, nzlat@cs.ntua.gr,\\
\small \tt pikrakis@unipi.gr, potam@central.ntua.gr}

\begin{document}
\ninept
\maketitle
\begin{abstract}
The recent development of Audio-based Distributional Semantic Models (ADSMs) enables the computation of audio and lexical vector representations in a joint acoustic-semantic space.
In this work, these joint representations are applied to the problem of automatic tag generation. 
The predicted tags together with their corresponding acoustic representation are exploited for the construction of acoustic-semantic clip embeddings. 
The proposed algorithms are evaluated on the task of similarity measurement between music clips. 
Acoustic-semantic models are shown to outperform the state-of-the-art for this task and produce high quality tags for audio/music clips.
\end{abstract}
\section{Introduction}
\label{sec:intro}
Semantic information in the form of metadata, e.g., tags, has been valuable in enhancing the performance for many music processing tasks \cite{levy2009music,lin2011exploiting,lamere2008social}.
Metadata typically comes in two forms:  free text associated with a music or audio clip or tags (list of words or phrases) that describe the clip. Tags are often preferred over (web-mined) text, because they give a direct description of the song, e.g., genre or instruments, while the latter is inherently noisy; only a part of the text is musically relevant.
The automatic annotation of clips (auto-tagging) is becoming vital and finds numerous applications including efficient music indexing and retrieval.

A variety of auto-tagging methods have been proposed in the literature.
In \cite{kim2009using}, the similarity between artists is exploited for the prediction of the most descriptive tags for clips. 
 In \cite{mandel2011contextual}, language models were computed using Restricted Boltzmann Machines, while in \cite{seyerlehner2011draft}, the combination of audio features is proposed within a block-level framework.
The use of semantic tags for music similarity measurement is proposed in \cite{turnbull2008semantic}, where each song is represented by a Semantic Multinomial Distribution over a vocabulary of tags.

Music similarity is at the core of query-by-example, where the user gives a musical piece as a query and the system returns a ranked list of recommendations.
Content-based similarity can be exploited by collaborative filtering algorithms~\cite{mcfee2012learning} especially where there is lack of collaborative filtering data, a.k.a. ``cold start" problem.
This improves the efficiency of music recommendation and playlist generation, two important tasks in Music Information Retrieval (MIR).
Music similarity estimation can be formulated as 
the problem of finding an appropriate embedding of a music clip with respect to a distance metric.  
Many approaches
use machine learning techniques in order to learn the distance metric that best approximates absolute user ratings 
\cite{wolff2012systematic,stober2011experimental} 
or relative user ratings 
from different datasets~\cite{wolffcomparative}.
Furthermore, research efforts include the collection of similarity scores that can be used as groundtruth information~\cite{law2009evaluation,ellis2002quest} and the investigation of different evaluation techniques~\cite{logan2003toward} for music similarity.

Distributional Semantic Models (DSMs) \cite{baroni2010distributional} is a popular method for automatically constructing semantic representations from text.
Despite their success in various semantic tasks (e.g., semantic classification and computation of semantic similarity), the DSMs have been criticized as ``disembodied", since they rely solely on linguistic information 
without being grounded in perception and action.
 The disconnection of natural language from the physical world, also referred as the symbol grounding problem \cite{harnad1990symbol}, 
 is alleviated via the integration of multiple modalities \cite{bruni2011distributional,iosifcrossmodal}.
The development of audio-based DSMs (ADSMs) was proposed in \cite{lopopolo-vanmiltenburg:2015:IWCS2015} for the representation of words based on their acoustic properties, while in \cite{kiela-clark:2015:EMNLPa} an extension was presented using combinations of auditory and linguistic features. 
A recently proposed approach dealt  with the fusion of (different) acoustic features according to the nature of sounds (music, speech, other) \cite{karamanolakis2016audio}. 

In this work, the ADSM described in \cite{karamanolakis2016audio} is adopted for the computation of audio and lexical vector representations in a joint acoustic-semantic space. These `bag-of-audio-words' representations are used for the automatic annotation of music clips.
Then, the predicted tags and the acoustic features are exploited for the construction of acoustic-semantic clip embeddings.
The proposed algorithms are evaluated on the task of music similarity measurement between clips taken from the MagnaTagATune dataset \cite{law2009evaluation}.

This paper is organized as follows. In Section~\ref{sec:system}, various methods are described for the construction of clip representations.
In Section~\ref{s:eval-dataset}, the experimental dataset and procedure are described in detail, while
in Section~\ref{s:eval-results}, the evaluation results of the proposed methods are reported and compared with the literature.

\begin{figure*}
\begin{subfigure}{\textwidth}
  \centering
  \centerline{\includegraphics[scale=0.5]{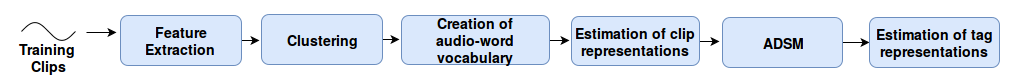}}
\caption{}
\vspace{-1.0mm}
\label{fig:train}
\end{subfigure}

\begin{subfigure}{\textwidth}
  \centering
  \centerline{\includegraphics[scale=0.5]{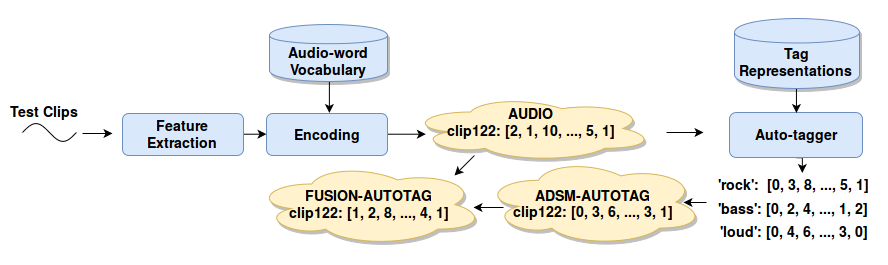}}
  \vspace{-4.0mm}
\caption{}

\label{fig:test}
\end{subfigure}
\caption{Schematic explanation of the proposed algorithms. (a) Training: creation of the audio-word vocabulary and tag representations.\\ (b) Testing: an auto-tagging example along with various methods for the representation of clips: AUDIO, ADSM/FUSION-AUTOTAG.}
\label{fig:system}
\end{figure*}

\section{System description}
\label{sec:system}
For a dataset consisting of audio clips and the corresponding annotation (tags), the procedure for the creation of clip representations consists of a series of steps described below.

\subsection{Creation of audio-word vocabulary}
\label{s:aw-vocabulary}
In order to train an audio-word vocabulary,  
the clips are partitioned into partially overlapping windows and a feature vector is extracted from every window.
Therefore, every clip is represented by a set of vectors depending
on its length. Next, all vectors are clustered by applying the $k$-means algorithm and the $k$ centroids of the returned clusters are considered as the audio-words of the vocabulary.

\subsection{Audio clip representations}
\label{ss:acoustic-space}
As presented in \cite{karamanolakis2016audio}, audio clips are represented as a mixture of audio-words.
For each window $\vec{o_t}$ of a clip, a
feature vector $\vec{x_t}\in R^d$ is computed (where $d$ is the dimensionality of the feature space)
and associated with the audio-word vocabulary.
The association is performed by encoding the $\vec{x_t}$ vector as a $k$-dimensional vector $\vec{e_t}$:
\begin{equation}
\label{eq:hard}
\vec{e_{t}} = (w_1, w_2, ..., w_k),
\end{equation}
where $w_i \in [0,1]$ and $\sum_{i=1}^{k}w_{i}=1$.
The weight $w_i$ denotes the contribution of the $i$-th audio-word to the window representation and is computed according to the similarity  score between the audio-word and the feature vector $x_t$ (see \cite{karamanolakis2016audio} for details).
Finally, the clip representation, $r_c$, is obtained by averaging the vectors computed for the respective windows.
Given a collection consisting of $M$ clips, this process results in a $M \times k$ matrix. 
The space consisting of bag-of-audio-words clip representations, will be referred as AUDIO space.

\subsection{Tag representations via the ADSM}
\label{ss:tag-representation}
As described in \cite{karamanolakis2016audio}, an ADSM is constructed that provides bag-of-audio-words embeddings for tags, based on their association with the clips.
In particular, the representation $\vec{r_j}$ of a tag $j$ is computed by averaging the representations of clips that have this tag in their descriptions.
For a collection of audio clips with $T$ (unique) tags this results in a $T \times k$ matrix.
Then, the Positive Pointwise Mutual Information (PPMI) weighting is applied to the matrix~\cite{bullinaria2007extracting}.
The overall procedure, up to the creation of tag representations, is represented in Fig~\ref{fig:system}(\subref{fig:train}).
\begin{table*}[!htb]
\centering
\begin{tabular}{|c|c|c|}
\hline
Clip id & Groundtruth Tags & Predicted Tags  ($N$=5)\\ \hline\hline
3843 &\textbf{indian}, \textbf{sitar} & \textbf{sitar}, \textbf{indian}, eastern, india, oriental \\ \hline

   9531 & rock, \textbf{heavy}, \textbf{heavy metal}, \textbf{loud}, fast, hard rock, \textbf{metal} &hard, \textbf{loud}, \textbf{heavy}, \textbf{heavy metal}, \textbf{metal} \\ \hline 

 13526 & bass, \textbf{drums}, drum, \textbf{funky}, \textbf{reggae} & \textbf{funky}, beat, \textbf{drums}, \textbf{reggae}, funk \\ \hline

15380 &\textbf{classical}, \textbf{solo}, \textbf{cello}, \textbf{violin}, strings & \textbf{cello}, viola, \textbf{violin}, \textbf{solo}, \textbf{classical} \\ \hline

19920 & - & orchestra, violins, flutes, fiddle, violin \\ \hline 
   21725 & choir, \textbf{choral}, \textbf{men}, man & monks, chant, chanting, \textbf{men}, \textbf{choral} \\ \hline

29231 & \textbf{acoustic}, \textbf{guitar} & classical guitar, \textbf{guitar}, \textbf{acoustic}, lute, spanish\\ \hline
 43390 & \textbf{rock}, loud, \textbf{pop}, vocals, \textbf{male vocals} & \textbf{male vocals}, \textbf{pop}, male vocal, male singer, \textbf{rock} \\ \hline

 48010 & silence & low, soft, no singing, quiet, wind \\ \hline
57081 &\textbf{piano} & piano solo, \textbf{piano}, classic, solo, classical \\ \hline
\end{tabular}
\caption{\small{ Examples of auto-tagging outputs for the MagnaTagATune dataset.}}
\label{table:auto-tagging-examples}
\end{table*}

\subsection{Clip annotation (auto-tagging)}
\label{ss:autotagging}
The bag-of-audio-words representations of tags provide a straightforward way for the automatic annotation (auto-tagging) of an audio clip $c$. First, the clip gets a representation, $\vec{r_c}$ in the AUDIO Space. 
Then, the cosine similarity is computed between $\vec{r_c}$ and the representation $\vec{r_j}$ of each tag $j$:
\begin{equation}
	s_{cj} = \frac{\vec{r_c} \cdot \vec{r_j}}{|\vec{r_c}| \cdot |\vec{r_j}|}.
\end{equation}
The $N$ tags that best describe the clip $c$ are those corresponding to the $N$ highest similarity scores $s_{cj}$.

Table~\ref{table:auto-tagging-examples} includes some examples selected from the MagnaTagATune dataset (see Section~\ref{s:eval-dataset}) for which the groundtruth labels are compared with the $N=5$ automatically predicted tags.
It is observed that many tags appear both in the predicted and the groundtruth labels, while other tags have very similar meaning (e.g., `silence' and `quiet').
Moreover, quite descriptive tags are returned for clips that have no annotations (e.g., clip 19920).

\subsection{Semantic representations of clips}
\label{ss:semantic-representation}
The ADSM described in Section~\ref{ss:tag-representation} is exploited for the representation of audio clips via their association with tags.
Specifically, the embedding of a clip is obtained by averaging the representations of tags that are semantically related with the clip.
The determination of clip-tag associations can be performed either using already provided metadata, 
or via the proposed auto-tagging scheme.
In the latter case, the clip gets a representation (in the ADSM-AUTOTAG space), as follows:
\begin{equation}
\vec{r_c'} = \frac{1}{N}\sum_{i=1}^{N} \vec{r_i},
\end{equation}
where $\vec{r_i}$ is the representation of the i-th tag, while $N$ denotes the number of tags returned by the auto-tagger.

\subsection{Fusion of audio and semantic representations}
\label{sss:semantic-acoustic-fusion}
The similarity score between two clips in the AUDIO space is computed as the cosine similarity of their acoustic representations.
Similarly, the similarity between clips in the ADSM space is computed as the cosine similarity of their respective semantic representations.
However, the characterization of clips only via their semantic representations in some cases may lead to inaccurate estimates.
For example, if two clips are annotated with the same labels, they will have exactly the same ADSM representations although they sound different.
This problem can be alleviated with the fusion of acoustic and semantic information.
The clip embedding $\vec{r_c}$ in the AUDIO space can be combined with the embedding from the ADSM (or ADSM-AUTOTAG) space, $\vec{r_c'}$, via a weighted average scheme: 
\begin{equation}
	\vec{r_c''} = w \vec{r_c'} + (1-w) \vec{r_c},
	\label{eq:weighted-average}
\end{equation}
or via a weighted concatenation:
\begin{equation}
	\vec{r_c''} = w \vec{r_c'} \oplus (1-w) \vec{r_c},		\label{eq:weighted-concatenation}
\end{equation}
where $\oplus$ stands for the vector-concatenate operator and $w$ is a real-valued number indicating the relative importance of the fused representations.
This method results in clip embeddings in the FUSION (or FUSION-AUTOTAG) space. 
The various methods for the representation of clips are graphically represented along with an auto-tagging example in Fig~\ref{fig:system}(\subref{fig:test}).

\section{experimental dataset and procedure}
The proposed algorithms are evaluated for the similarity computation between songs.
For this purpose, the MagnaTagATune dataset is used.

\subsection{The MagnaTagATune dataset}
\label{s:eval-dataset}
The MagnaTagATune dataset contains 25,863 30-second audio clips (provided by the Magnatune\footnote{http://magnatune.com/} label) and 188 tags.
In addition, similarity data have been collected from the TagATune game \cite{law2009input}. 
In a bonus part of the game, the user listens to three songs and gives a vote to the song that sounds as the most dissimilar when compared with the other two songs (often called as odd one out game). 
The similarity data for every triplet of clips is stored in form of a triplet representing the histogram of votes i.e., the clip associated with the maximum value is the most irrelevant clip. 
In total, 533 triplets are derived and every triplet of song ids is saved in the form $(a,b,c)$ where $c$ is the outlier. 
This explicitly means that $d(a,b)<d(a,c)$ (constraint $(a,b,c)$) and $d(b,a)<d(b,c)$ (constraint $(b,a,c)$), where $d()$ is the perceptual distance between two clips.

Due to the objective opinion of each user, there are constraints that contradict each other. Hence, a method was proposed in \cite{stober2011experimental} in order to deal with the inconsistent constraints. In \cite{wolff2012systematic}, the 860 remaining constraints derived from \cite{stober2011experimental} were split into non-overlapping training and test sets of 774 and 86 constraints, respectively enabling 10-fold cross-validation. 
These constraints are published 
serving as a common evaluation benchmark.
The described benchmark is used in this work.

\subsection{Data preprocessing and feature extraction}
\label{ss:preprocessing}
All audio clips are converted to WAV format and resampled at 22.05 kHz.
For each clip, a feature vector is extracted from windows of $250$ ms with a step of $100$ ms.
Here, two different feature vectors are extracted: the first (EchoNest) is obtained by the EchoNest API 1.0\footnote{http://developer.echonest.com/}. 
Specifically, 24 features per audio frame have been kept, consisting of 12 chroma features and 12 timbre features. The chroma features describe the relative dominance of every pitch in the chromatic scale and are normalized to $[0,1]$. The timbre features correspond to the coefficients of 12 basis functions which represent the texture of sound.
The second type of feature vectors (MFCCdd) consists of the Mel Frequency Cepstral Coefficients (MFCCs) (concatenated with spectral energy), their 1st and 2nd order derivatives, resulting in a vector of 39 coefficients.
In both cases, the feature vectors are normalized by their mean and standard deviation values (Z-normalization).
\subsection{Experimental procedure}
For each step of the 10-fold cross validation, the audio-word vocabulary and the ADSM are built using the training clips\footnote{For computational efficiency, 1000 clips are randomly selected from the training set for the audio-word vocabulary.}.
The representations are then computed for the test clips and the similarity scores are obtained, as described in Section~\ref{sss:semantic-acoustic-fusion}.
 The following methods are evaluated:
\begin{itemize}
\item \textbf{AUDIO:} only the acoustic features are used (Section~2.3).
\item \textbf{ADSM:} clip representations are obtained via the provided tags (Section~\ref{ss:semantic-representation}).
\item \textbf{ADSM-AUTOTAG:} same with ADSM method, but all clip representations (even for clips that have labels) are derived via auto-tagging (Section~\ref{ss:autotagging}). Here, $N=20$ tags are predicted for every clip.
\item \textbf{FUSION:} fusion of AUDIO and ADSM representations (Section~\ref{sss:semantic-acoustic-fusion}, for $w=0.9$).
\item \textbf{FUSION-AUTOTAG}: fusion of AUDIO and ADSM-AUTO-TAG representations, for $w\!=\!0.9$ and $N\!=\!20$.
\end{itemize}
The number of audio-words (i.e., the number of clusters) is fixed to $k=300$. 
In addition, dimensionality reduction via Singular Value Decomposition ($svd$ denotes the number of dimensions) was optionally performed on the matrix where the rows correspond to different audio clips and the columns to their representations. 

The adopted evaluation metric is the accuracy of each method, which is defined as the percentage of total test constraints (see Section~\ref{s:eval-dataset}) that are satisfied. 
The experimental procedure is followed for each of the 10 folds and the final score is computed as the average of the accuracy scores.
This procedure is repeated 10 times.

\begin{table}[t]
\centering
\begin{tabular}{|c|c|}
\hline
\textbf{Literature Method}&\textbf{EchoNest Features}\\ \hline
Euclidean [11] &0.598 \\ \hline
RITML [11] & 0.711\\ \hline
SVM [9] & \textbf{0.712} \\ \hline
MLR [9] & 0.689 \\ \hline
\end{tabular}
\caption{\small{Accuracy of methods reported in the literature \cite{wolff2012systematic,wolffcomparative}.}}
\vspace{3mm}
\label{table:literature-results}
\end{table}

\begin{table}[t]
\centering
\begin{tabular}{|c|c|c c|c c|}
\hline
\multicolumn{2}{|c|}{\textbf{Proposed}} & \multicolumn{2}{|c|}{\textbf{EchoNest}} &\multicolumn{2}{|c|}{\textbf{MFCCdd}}\\ 
\multicolumn{2}{|c|}{\textbf{Method}} & $k$=300 & $svd$=10 & $k$=300& $svd$=10 \\ \hline
\multicolumn{2}{|c|}{AUDIO} & 0.613  &0.644& 0.636 &0.646\\ \hline
\multicolumn{2}{|c|}{ADSM} & 0.705 &0.719&  0.717 &0.720\\ \hline
\multicolumn{2}{|c|}{FUSION} & 0.720 &\textbf{0.731} & 0.681 &0.684 \\ \hline
\multicolumn{2}{|c|}{ADSM-AUTOTAG} & 0.705 &0.705 &  0.693 &0.696\\ \hline
\multicolumn{2}{|c|}{FUSION-AUTOTAG} &  0.705 &0.709 & 0.662 &0.672 \\ \hline
\end{tabular}
\vspace{1mm}
\caption{\small{ Accuracy of proposed methods for EchoNest and MFCCdd features. For FUSION, $w$=0.9, while for AUTOTAG, $N$=20.}}
\label{tab:results}
\end{table}


\section{Evaluation results}
\label{s:eval-results}
Table~\ref{table:literature-results} includes the state-of-the-art performance (see \cite{wolffcomparative} for a brief overview), while
the accuracy of the proposed methods is presented in Table~\ref{tab:results}.
In addition to clip representations, where $k$=300, the performance is reported with respect to $svd$=10 dimensions where the best accuracy is achieved for most methods.
The FUSION method\footnote{The performance of the weighted concatenation (see (\ref{eq:weighted-concatenation})) was found to be comparable with the performance of the weighted average (see (\ref{eq:weighted-average})), so the respective results are not reported here.} yields the best accuracy score: 0.731, which is higher than the best score reported in the literature (0.712 achieved by SVM \cite{wolff2012systematic}).
Moreover, the exploitation of semantic information via the  ADSM method boosts the performance compared with the AUDIO method (up to 12.7\% relative improvement).
Regarding the ADSM-AUTOTAG, the auto-tagging algorithm is applied for all the test clips, which were not used for the training representations. However, the ADSM-AUTOTAG method achieves comparable performance with the ADSM method, where the dataset labels are used.

The quality of the auto-tagger's predictions was also confirmed after the manual inspection of the predicted tags for the examples of Table~1.
Hence, the auto-tagger's predictions can be exploited for the annotation of clips without tags or for the enrichment of provided annotations. 
Despite the fact that the training similarity data (here, in the form of constraints) are not used, the reported results exceed the state-of-the-art. 
\begin{figure}[t]
  	\centering
  	\includegraphics[height=6.0cm, width=8.5cm]{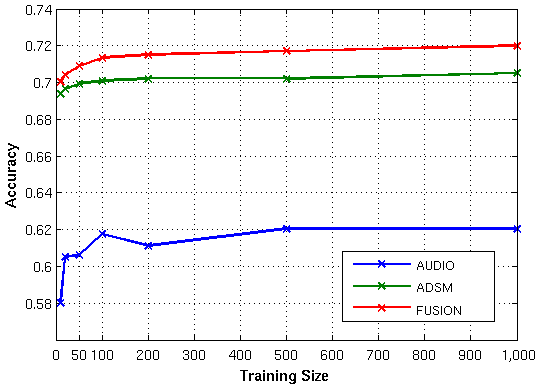}
  	\vspace{1.0mm}
    \caption{Accuracy with respect to different training sizes, i.e., number of MagnaTagATune clips used to train the audio-word vocabulary. Results are shown for EchoNest features using $k=300$.}
  	\label{fig:test-train-size}
    \vspace{-3.0mm}
\end{figure}
Hence, the proposed unsupervised algorithm can be applied in any dataset without the need of manually created similarity data.

In Fig~\ref{fig:test-train-size}, the accuracy of the AUDIO, ADSM and FUSION methods is shown as a function of the number of clips, which were used for the construction of the audio-word vocabulary. 
Interestingly, even a small number of clips (50-200) is sufficient for the creation of a vocabulary of audio-words, achieving 0.717 accuracy (for FUSION method and 200 training clips). 
In addition, it appears that the ADSM and FUSION methods are more robust with respect to data sparsity compared to the AUDIO method.

\section{Conclusions}
In this work, an audio-based DSM is exploited for creating lexical representations and the clip representations are derived via an auto-tagging scheme.
The evaluation is performed on the task of music similarity measurement using the MagnaTagATune dataset.
Although 
the similarity ratings provided with the dataset 
are not considered in the proposed algorithms,
they exceed state-of-the-art methods, that use them in order to train a similarity metric.
Moreover, very few data, e.g., 50 clips, are adequate to train the vocabulary, which is used for the creation of bag-of-audio-words representations.
Therefore, this unsupervised algorithm can be applied in every dataset where neither clip annotations or similarity data are necessary.
Regarding future work, we aim to investigate the
fusion of acoustic, semantic and visual modalities and test the proposed algorithm on multimedia data. 
\section{Acknowledgements}
This work has been partially supported by the BabyRobot
 project supported by EU H2020 (grant \# 687831).

\newpage
\vfill\pagebreak

\bibliographystyle{IEEE}

\bibliography{strings,references}
\end{document}